\documentclass[runningheads]{llncs}
\usepackage[T1]{fontenc}
\usepackage{graphicx}
\usepackage{hyperref}
\usepackage[style=authoryear,sortcites=true]{biblatex}
\usepackage{changepage}
\usepackage{amsfonts}
\usepackage{amsmath}
\usepackage{xcolor}
\usepackage[ruled,vlined]{algorithm2e}

\definecolor{customcite}{HTML}{006896}
\hypersetup{
  colorlinks   = true,
  urlcolor     = customcite,
  linkcolor    = customcite,
  citecolor    = customcite, 
}
\usepackage{cleveref}
\DeclareCiteCommand{\plaincite}
  {\usebibmacro{prenote}}
  {\printtext[bibhyperref]{%
    \printnames{labelname}%
    \setunit{\printdelim{nameyeardelim}}%
    \printfield{year}}}
  {\multicitedelim}
  {\usebibmacro{postnote}}

\DeclareCiteCommand{\parencite}[\mkbibparens]
  {\usebibmacro{prenote}}
  {\printtext[bibhyperref]{%
    \printnames{labelname}%
    \setunit{\printdelim{nameyeardelim}}%
    \printfield{year}}}
  {\multicitedelim}
  {\usebibmacro{postnote}}

\bibliography{references.bib}
\graphicspath{ {Images/} }

\begin{document}
\title{Connectivity-Preserving Cortical Surface Tetrahedralization}
\author{Besm Osman\inst{1} \and
Ruben Vink\inst{1}\and
Andrei Jalba\inst{1}\and
Maxime Chamberland\inst{1}
}

\authorrunning{B. Osman et al.}
\institute{1. Department of Mathematics and Computer Science, Eindhoven University of Technology, Eindhoven, The Netherlands
}
\maketitle              % typeset the header of the contribution
\begin{abstract}
A prerequisite for many biomechanical simulation techniques is discretizing a bounded volume into a tetrahedral mesh. In certain contexts, such as cortical surface simulations, preserving input surface connectivity is critical. However, automated surface extraction often yields meshes containing self-intersections, small holes, and faulty geometry, which prevents existing constrained and unconstrained meshers from preserving this connectivity. We address this issue by developing a novel tetrahedralization method that maintains input surface connectivity in the presence of such defects. We also present a metric to quantify the preservation of surface connectivity and demonstrate that our method correctly maintains connectivity compared to existing solutions.

% \keywords{First keyword  \and Second keyword \and Another keyword.}
\end{abstract}

\section{Introduction}
Tetrahedralization is the process of decomposing a bounded volume, typically defined by a surface mesh, into discrete tetrahedra. This step is often required for computational simulations in biomechanics. When tetrahedra are uniform in size, computational simulations tend to be more accurate and stable \parencite{taniguchiUseDelaunayTriangulation2000}. Therefore, Delaunay tetrahedralization is often preferred, as it ensures that no point lies inside the circumsphere of any tetrahedron, avoiding poorly shaped elements that can lead to numerical instability. Specifically, constrained tetrahedralization, which guarantees that all faces from the input surface remain in the final result, is generally desired. \\

However, constrained tetrahedralization is not achievable for all surfaces. Constrained methods, such as TetGen \parencite{siTetgenDelaunayBasedQuality2022} and Gmsh \parencite{geuzaineGmsh3DFinite2009}, require a Piecewise Linear Complex (PLC) as input. This input must be free from geometric inconsistencies, such as self-intersections and topological errors that render the mesh non-manifold. While mesh repair tools exist to detect problematic regions and attempt to heal them, these tools may delete large portions of the mesh or merge self-intersecting regions, effectively fusing regions that are meaningfully distinct \parencite{atteneDirectRepairSelfintersecting2014,campenExactRobustSelfIntersections2010}. \\

This limitation is particularly acute in the context of cortical surface meshing. The human brain is characterized by a highly folded geometry, where deep folds (sulci) and ridges (gyri) are packed tightly together. To generate surface meshes for these structures, standard pipelines like FreeSurfer \parencite{fischlFreeSurfer2012} and BrainSuite \parencite{shattuckBrainSuiteAutomatedCortical2002} are used. These tools classify tissue from MRI scans to construct an inner white-gray matter interface, which is then `pushed' outwards along its normals to map the outer pial surface \parencite{eskildsenQuantitativeComparisonTwo2007}.\\

While this method captures anatomical structure with high accuracy, the expansion process frequently generates self-intersections within the narrow spaces of the cortical folds. Furthermore, mesh simplification is often required for computational efficiency, which increases the prevalence of these intersections. Different software packages also use inconsistent methods for self-intersection detection with varying floating-point tolerances; consequently, constrained meshing software may report self-intersections on surfaces seen as `intersection-free' by other tools. Due to these geometric issues, generating constrained volumetric models of cortical surfaces is difficult and error-prone.\\

Alternatively, to circumvent the strict requirements of constrained methods, researchers often use unconstrained tetrahedralization. These methods do not require the input to be free from self-intersections, but they also do not guarantee that the output will retain the input surface faces. Existing neuroanatomical meshing pipelines use variations of these methods to allow for finite element simulations \parencite{tranImprovingModelbasedFunctional2020, fangTetrahedralMeshGeneration2009, ledermanGENERATIONTETRAHEDRALMESH2011}. Recent general meshing methods like TetWild \parencite{huFastTetrahedralMeshing2020} aim to preserve most input surface faces while remaining robust to faulty input geometry, targeting an almost-Delaunay tetrahedralization rather than a strictly constrained one. However, these approaches do not preserve the anatomical connectivity of the cortical folds. For instance, if a cortical surface contains two gyri (convex hills) that self-intersect due to extraction inaccuracies, these methods often create erroneous connections between them. These artificial fusions drastically reduce accuracy in both deformation and non-deformation studies, limiting the feasibility of subject-specific volumetric cortical fold studies. \\

In this study, we present a novel tetrahedralization method that maintains the surface connectivity of cortical surfaces, even in the presence of self-intersections. To validate our approach, we introduce a metric to quantify how well the input geometry's connectivity is preserved in the volumetric mesh. We evaluate our method quantitatively using this metric and qualitatively via a deformation simulation compared against a ground truth.

\newpage
\section{Method} \label{section-tet-method}
Our method takes as input a cortical surface represented by a mesh consisting of vertices with spatial positions, connected by triangular faces. While intended to represent a manifold, the surface may include faulty geometry such as small holes and self-intersections. \Cref{section-algorithm} outlines the tetrahedralization algorithm, followed by implementation details in \cref{section-implementation}.  Given the construction method of the input cortical surface meshes \parencite{shattuckBrainSuiteAutomatedCortical2002}, we can assume that the connectivity of the mesh is correct, and we try to preserve this as much as possible during tetrahedralization. In \cref{section-connectivity-metric} we define this connectivity distance concretly and a metric to measure the preservation of this metric.  Finally, \cref{section-evaluation} details our dataset and evaluation methodology.

\subsection{Connectivity-Preserving Tetrahedralization Algorithm}\label{section-algorithm}

The proposed algorithm performs tetrahedralization while maintaining surface connectivity distance. The algorithm can deal with self-intersecting regions and malformed faces. It works for both 2D triangulations and 3D tetrahedralizations. Here we describe the 3D case, as the 2D case is analogous. A high-level outline of the algorithm is as follows:

\begin{enumerate}
\item Initialize the point set $\mathbf{P}$ by duplicating each surface vertex and offsetting the copies along a non-degenerate direction (e.g., the vertex normal) by positive and negative $\epsilon$.
\item Add internal points to $\mathbf{P}$, ensuring that the minimal distance between any internal point and any twin surface point in $\mathbf{P}$ is at least $2\epsilon$.
\item Generate a Delaunay tetrahedralization of $\mathbf{P}$.
\item Remove all tetrahedra that intersect any surface face of the original input mesh.
\item Find the main connected component, removing non-connected tetrahedra and points.
\item Move the twin surface points back to their original positions.
\end{enumerate}

\begin{figure*}[t]
  \includegraphics[width=\linewidth]{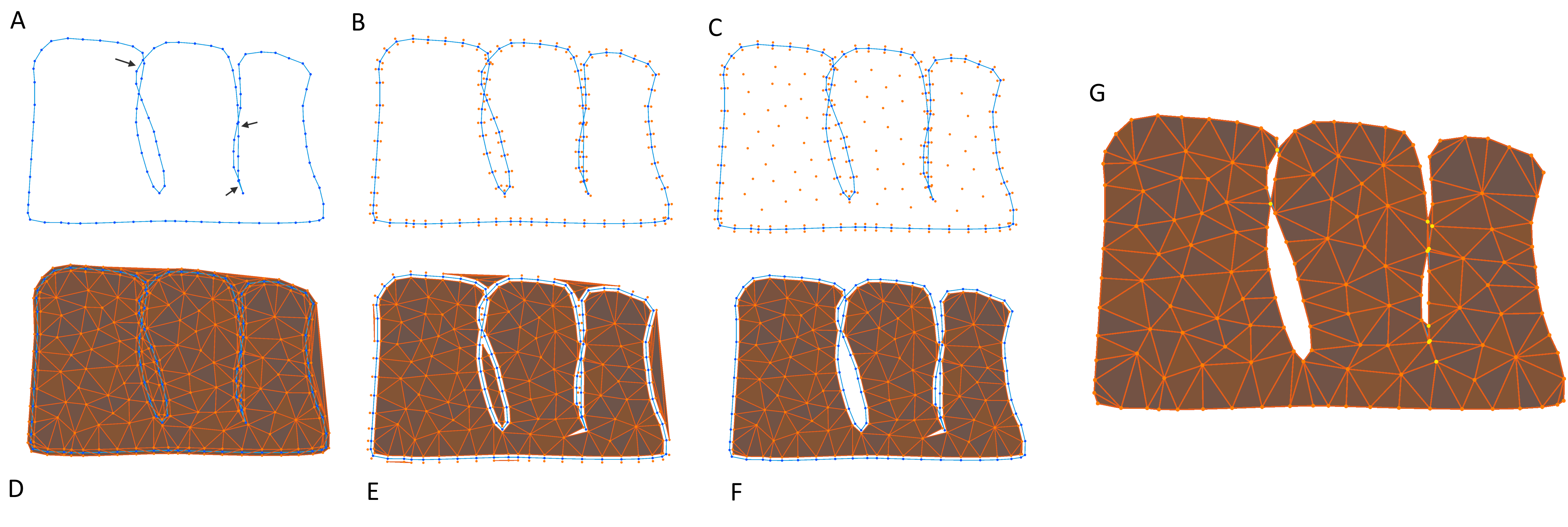}
  \caption{Illustration of the 2D version of our algorithm on a mesh with self-intersecting regions. (\textbf{A}) Input surface with overlapping regions indicated by arrows. (\textbf{B}) Construction of the point set $\mathbf{P}$ with twin points determined by node normal directions. (\textbf{C}) Addition of approximately uniform internal points to $\mathbf{P}$. (\textbf{D}) Triangulation of the point set $\mathbf{P}$. (\textbf{E}) Removal of any tetrahedra (or triangles in 2D) intersecting the original faces. (\textbf{F}) Elimination of disconnected components, keeping only the main connected component. (\textbf{G}) Restoration of twin nodes to their original positions, yielding the final mesh. Input nodes and edges are colored blue, triangulation edges are orange, and twin nodes are highlighted in yellow.} \label{fig-2d}
\end{figure*}

The algorithm starts by building a set of points $\mathbf{P}$ that will be tetrahedralized. This set $\mathbf{P}$ is initialized by duplicating each vertex of the surface, with the two copies offset in opposite (positive and negative) non-degenerate directions by a small value $\epsilon$, which ideally is smaller than the minimum distance between any pair of surface nodes (default $10^{-9}$ cm). The chosen direction should be non-degenerate with respect to all faces connected to the vertex. This construction ensures that the original surface lies within a thin layer defined by these twin points, a property utilized later in the process. For any original vertex, its closest point will be its own twin. Consequently, a direct connection between twins is necessarily formed in the Delaunay tetrahedralization \parencite{mausAllClosestNeighbors2010}. Consequently, any tetrahedra generated during the Delaunay step that bridge across the original surface tend to be confined to this thin, sliver-like region between the twin points. The choice of offset direction does not affect the connectivity of the final mesh but does influence the quality of the tetrahedral structure at the surface. Sophisticated methods can be used to choose a direction that maximally separates regions; however, through testing, we found that simply taking the vertex-based normal direction as the split direction works well in practice. \\

At this stage, internal points can be added to $\mathbf{P}$. Any set of points can be used (including the empty set), as long as the minimal distance between any internal point and any twin surface point in $\mathbf{P}$ is larger than $2 \epsilon$. This minimal distance reduces the number of non-conforming faces and makes the algorithm more robust to small holes. Points do not \emph{need} to be internal; however, tetrahedron quality near the surface is improved if external points and points in self-intersecting regions are avoided at this stage. For surfaces with holes, there is no true boundary separating internal from external, and detecting self intersecting regions  to exclude can be tedious. In \cref{section-implementation} we describe an iterative approach that adds semi-uniformly distributed internal points.  We specify the average distance between points to be equal to the average surface edge length for optimal tetrahedron quality. Next, we tetrahedralize $\mathbf{P}$ using a standard Delaunay point cloud tetrahedralization algorithm, \emph{without} regard for the original surface faces. \\

The resulting tetrahedralization encloses the original surface from all sides with short, sliver-like tetrahedra near the original faces (see \cref{fig-2d}(D)). we can remove all tetrahedra that intersect the original surface faces. This effectively removes the thin layer of tetrahedra bridging the twin points across the original surface. Since the tetrahedra connecting a set of twin points necessarily span the location of the original surface face, they will be removed by this cutting step. Although the input surface may have defects, in general, the surface is locally Euclidean. Thus, removing these surface-intersecting tetrahedra splits the mesh into distinct connected components. Crucially, even where the surface self-intersects, the twin points lie on either side of the intersecting faces. The cutting step, therefore, separates these geometrically overlapping but topologically distinct regions. \\

The main connected component remaining after the cutting step represents the desired tetrahedralized volume. We find the main connected component using a flood fill algorithm starting from a known interior point, removing the non-connected tetrahedra and non-connected points. Any node known to be within the interior volume (e.g., far from the surface boundary) can serve as the starting point for the flood fill. For instance, a node near the center of the cerebral hemisphere, far from any problematic folding regions, would be suitable. Finally, the remaining offset surface points are moved back to their original positions. Although inversion is possible, the offset $\epsilon$ is much smaller compared to the tetrahedra feature size, making it highly unlikely in practice. Both twin points of an original vertex may exist in the final result, each connecting to different tetrahedra. These could potentially be merged when they are close in terms of connectivity distance, however, keeping both points can better express the connectivity near self-intersecting regions. This process yields a tetrahedralization that approximates a constrained result by preserving surface features, while crucially maintaining the input surface's connectivity, even in the presence of self-intersections.

\subsection{Implementation} \label{section-implementation}
Both 2D and 3D versions of the algorithm are implemented in NeuroTrace \parencite{osmanVoxlinesStreamlineTransparency2023}, an open-source C\# tractography and cortical fiber deformation project. To optimize several steps, the 3D version utilizes a voxel grid as a spatial partitioner. The bounding volume of the input surface is divided into voxels, typically 1 mm$^3$ in size. Each voxel maintains collections of references to vertices, input surface faces, and tetrahedra whose bounding boxes intersect the voxel. This structure allows efficient spatial queries, such as nearest neighbor searches and ray-surface intersection tests. \\

Internal points are generated iteratively. First, a random candidate position is generated within the bounding box of the input surface. Each candidate position is then checked to determine if it lies inside or outside the surface volume. To perform this check in the presence of self-intersections or small holes in the input surface, six rays are cast from the candidate position along the positive and negative axis-aligned directions (+X, -X, +Y, -Y, +Z, -Z). If a ray intersects the surface and its direction is generally aligned with the surface normal at the intersection point (i.e., ray direction $\cdot$ surface normal $>$ 0), the test for that specific ray direction considers the point \emph{inside}. If the ray does not intersect the surface, or if the ray direction and surface normal are not aligned (ray direction $\cdot$ surface normal $\leq$ 0), the test for that direction considers the point \emph{outside}. The candidate point is classified as inside or outside based on a majority vote over the results from the six ray directions. \\

To improve performance, another optimization is used. A voxel containing no surface faces must lie entirely inside or entirely outside the original surface. Therefore, the inside/outside status can be precomputed for all such ``empty'' voxels, minimizing the number of required raycasts. The computationally expensive six-ray test is performed only for candidate points located within voxels that contain surface faces (i.e., boundary voxels). To encourage a uniform distribution of internal points, each valid inside candidate is also subjected to a uniformity test. This check ensures that the candidate is at least a minimum distance $r$ from all points in the set $\mathbf{P}$. By default, $r$ is set to the average edge length of the input surface mesh. If a candidate point passes both the inside test and the uniformity test, it is added to the point set $\mathbf{P}$. The iterative process of generating interior points terminates when a predefined number of consecutive candidates fail the uniformity check, suggesting that the space is filled according to the chosen $r$. \\

The Delaunay triangulation of the final point set $\mathbf{P}$ is computed using a custom incremental Boyer-Watson implementation for the 2D case and the MIConvexHull library \parencite{mattcampbellMIConvexHull2025} for the 3D case. After triangulation, checks for overlaps between the generated tetrahedra and the input surface faces are performed efficiently using the voxel grid. Recall that each voxel stores references to the surface faces and tetrahedra whose bounding boxes intersect it. Therefore, checking for intersections only between tetrahedra and surface triangles stored within the same voxel is sufficient to identify all tetrahedra that improperly overlap or pierce the input surface. After these overlapping tetrahedra are removed, a vertex-based flood fill is performed on the remaining tetrahedral mesh. The flood fill starts from a node known to be within the volume (e.g., an interior point that is not in a self-intersecting region) to identify the main connected component of tetrahedra representing the interior region. If determining a known interior point is difficult, multiple seeds can be tested to find the largest component. As a final step, each twin point is moved back to its original position, removing $\epsilon$ offsets.

\subsection{Connectivity Metric}\label{section-connectivity-metric}
To assess the connectivity between folds, we first observe that erroneous tetrahedra are problematic when they connect different folds, effectively creating a shortcut that reduces the geodesic distance between these regions. The more the shortest path distance along the mesh surface between two points is reduced compared to the original surface, the larger the effect it has on subsequent deformation simulations. The opposite can also occur, where missing links in the volumetric mesh cause a single anatomical region to exhibit vastly different behavior. Using this observation, we define a metric to penalize such shortcuts introduced during tetrahedralization. \\

Let $\mathcal{V} = \{v_1, v_2, \ldots, v_m\}$ be the set of vertices of the input surface mesh (the reference). Let $\mathcal{T} = \{t_1, t_2, \ldots, t_n\}$ be the set of vertices on the surface of the generated tetrahedral mesh. The surface of the tetrahedral mesh is identified by collecting all triangular faces that belong to exactly one tetrahedron in the mesh. \\

To evaluate changes in surface connectivity, we compute shortest path distances between the tetrahedral surface vertices and a set of landmark positions $\mathcal{L} = \{l_1, l_2, \ldots, l_k\}$. These landmarks are chosen from the reference surface vertices $\mathcal{V}$.  Ideally, one would compare the surface path distances between all pairs of nodes on the tetrahedral mesh surface against the corresponding distances on the reference surface. However, this is computationally expensive, especially for large meshes. Therefore, we specify landmarks to allow for a subset of comparisons, allowing a tunable trade-off between accuracy and performance. For each tetrahedral surface vertex $t \in \mathcal{T}$ and landmark $l \in \mathcal{L}$, we compute $d_{\text{mesh}}(t, l)$ as the shortest path distance from $t$ to $l$ traveling only along the edges of the tetrahedral mesh's surface. Similarly, for each reference vertex $v \in \mathcal{V}$ and landmark $l \in \mathcal{L}$, we compute $d_{\text{ref}}(v, l)$ as the shortest path distance from $v$ to $l$ across the edges of the reference mesh. For every vertex $t$ on the tetrahedral surface, we compare its surface path distance to each landmark $l$ with the corresponding reference surface distance. Our tetrahedralization allows for direct mapping between vertices $\mathcal{T}$ and the vertices $\mathcal{V}$ of the reference surface. However, other meshing methods do not. To find the corresponding node on the reference surface, we identify the closest vertex on the reference surface for a given $t$. Let $\mathtt{closest}(t)$ be the vertex $v \in \mathcal{V}$ that is nearest to $t \in \mathcal{T}$ in terms of Euclidean distance. The discrepancy for a single landmark $l$ is then:
$$C(t, l) = |d_{\text{mesh}}(t, l) - d_{\text{ref}}(\text{closest}(t), l)| $$

To combine the discrepancies across all landmarks robustly, we define our connectivity metric $C(t)$ for each tetrahedral surface vertex $t$ as the median difference:
$$C(t) = \operatorname{median}_{l \in \mathcal{L}}(C(t, l)) = \operatorname{median}_{l \in \mathcal{L}}(|d_{\text{mesh}}(t, l) - d_{\text{ref}}(\text{closest}(t), l)|)$$

Ideally, $C(t)$ should be close to $0$, indicating that the local surface connectivity around $t$ is well-preserved compared to the reference. By taking the absolute value, this metric effectively identifies both inappropriate connections (shortcuts) between folds and missing connections that increase the geodesic distance. A high $C(t)$ value indicates that the path distances to landmarks along the tetrahedral surface are significantly different from the expected distances on the reference surface, highlighting problematic areas. A low score suggests that connectivity is properly maintained. A potential issue arises because $\mathtt{closest}(t)$ uses Euclidean distance. If the input surface has folds that are spatially close but distant along the surface, the $\mathtt{closest}(t)$ mapping might associate $t$ with a reference vertex $v$ on a different fold. However, occurrences are rare. By taking the median of the discrepancies over multiple landmarks in $\mathcal{L}$, the metric is robust toward these potential outliers.

\subsection{Evaluation} \label{section-evaluation}
To validate our method, we generated a cortical pial mesh from an HCP Young Adult subject \parencite{vanessenWUMinnHumanConnectome2013} using the BrainSuite cortical surface extraction pipeline \parencite{shattuckBrainSuiteAutomatedCortical2002}. The vertex count was reduced, and the mesh was subsequently post-processed using a smoothing filter in Blender \parencite{blendercommunityBlender3DModelling}. The resulting surface comprised 614,548 nodes and 1,229,092 triangular faces. This surface contains thousands of self-intersecting faces both before and after processing. The exact count of self-intersections is difficult to determine due to the use of inexact floating-point arithmetic and varying detection tolerances or thresholds across different software packages. For example, Gmsh, TetGen, Blender, and our custom code all reported different numbers of self-intersections, typically around 10,000 intersecting faces. \\

Since Gmsh and TetGen were unable to mesh this surface due to the intersections, we compared our results against two popular techniques capable of handling self-intersections: Isosurface Stuffing \parencite{labelleIsosurfaceStuffingFast2007} and TetWild \parencite{huTetrahedralMeshingWild2018}. We ran TetWild using the official Docker container with its default arguments. For Isosurface Stuffing, we used the implementation provided in QTetramesher \parencite{lubkeQtetramesher2025}. We decreased the facet size parameter in QTetramesher to generate a tetrahedral mesh with a count comparable to that produced by TetWild. The resulting number of tetrahedra was 954,180 for Isosurface Stuffing and 935,484 for TetWild. We ran our algorithm as previously described, resulting in a mesh containing 812,641 tetrahedra. All meshes have their units set to meters.

\newpage
\section{Results}\label{section-tet-results}

\begin{figure*}[t]
    \centering
    \includegraphics[width=\linewidth]{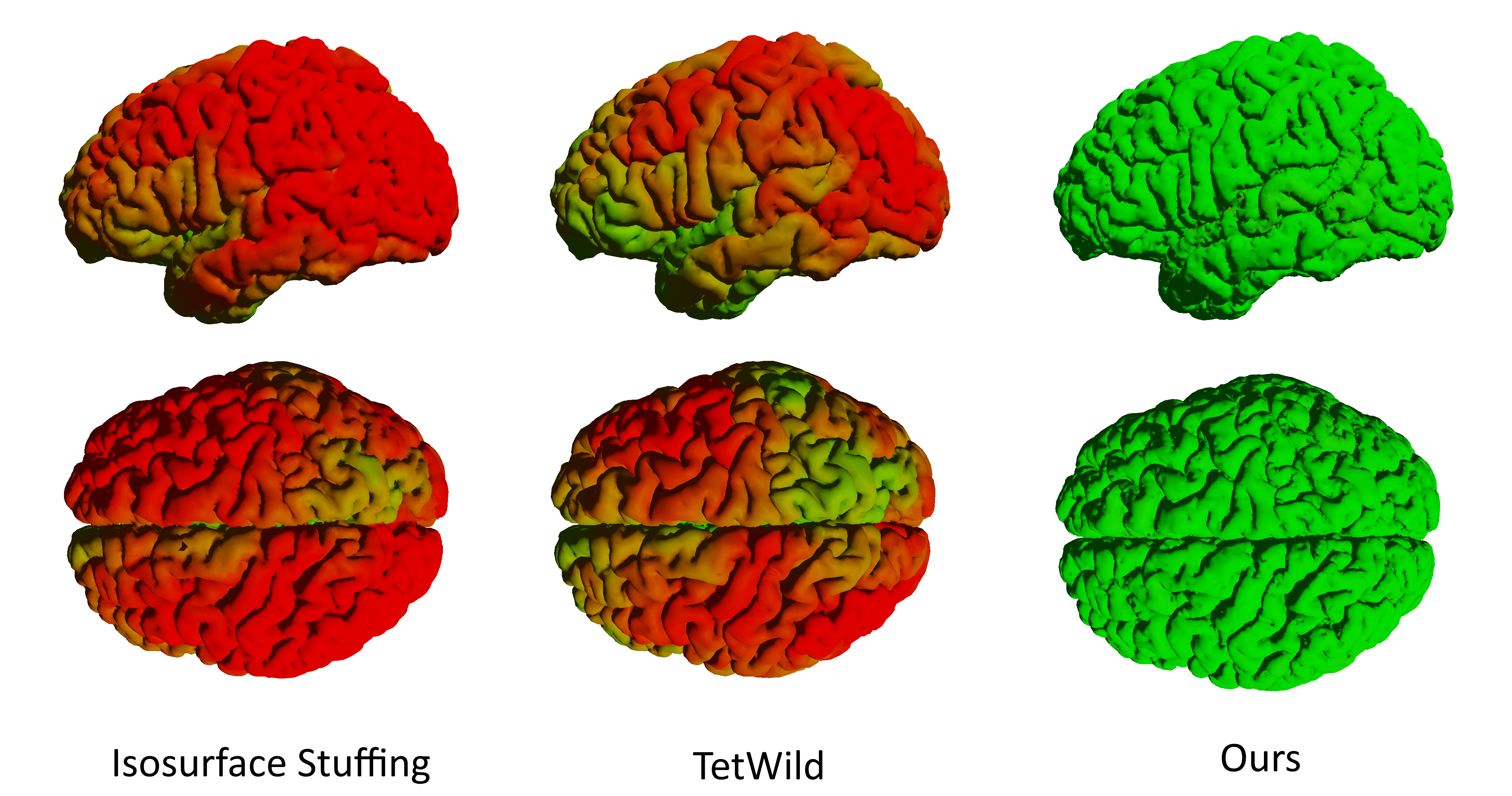}
    \caption{Comparison of tetrahedralized mesh surface connectivity. Each surface vertex $v$ is colored by linear interpolation between green (low discrepancy) and red (high discrepancy), with the interpolation factor determined by the connectivity metric $C(v)$ described in \cref{section-tet-method}. Green indicates good agreement between the reference surface connectivity and the tetrahedral mesh surface connectivity, while red indicates significant differences.}
    \label{fig-connectivity-comp}
\end{figure*}

\subsection{Connectivity Evaluation} 
We compute our previously described connectivity metric with $|\mathcal{L}| = 32$ landmarks, populated with vertices placed at regular intervals on the input surface. To compute this metric, we first precompute the approximate geodesic distances from each landmark to all vertices on both the tetrahedral mesh and the reference surface using Dijkstra's algorithm. Using a spatial partitioner, we then find the nearest neighbor pairings between the vertices of the tetrahedral mesh and the vertices of the reference surface. In \cref{fig-connectivity-comp}, we visually present the resulting connectivity score for each vertex. Note how the mesh generated by Isosurface Stuffing exhibits the most significant errors (indicated by red regions), while TetWild performs better overall but still suffers from similar connectivity issues. The median connectivity score $C(v)$ (where lower values indicate better connectivity) is $4.76 \times 10^{-2}$ for the Isosurface Stuffing mesh, $3.43 \times 10^{-2}$ for the TetWild mesh, and $0.45 \times 10^{-2}$ for the mesh generated by our method.

\subsection{Smoothing Simulation} 

\begin{figure*}[t] 
  \begin{adjustwidth}{-.2cm}{-.2cm} 
    \centering
    \includegraphics[width=\linewidth]{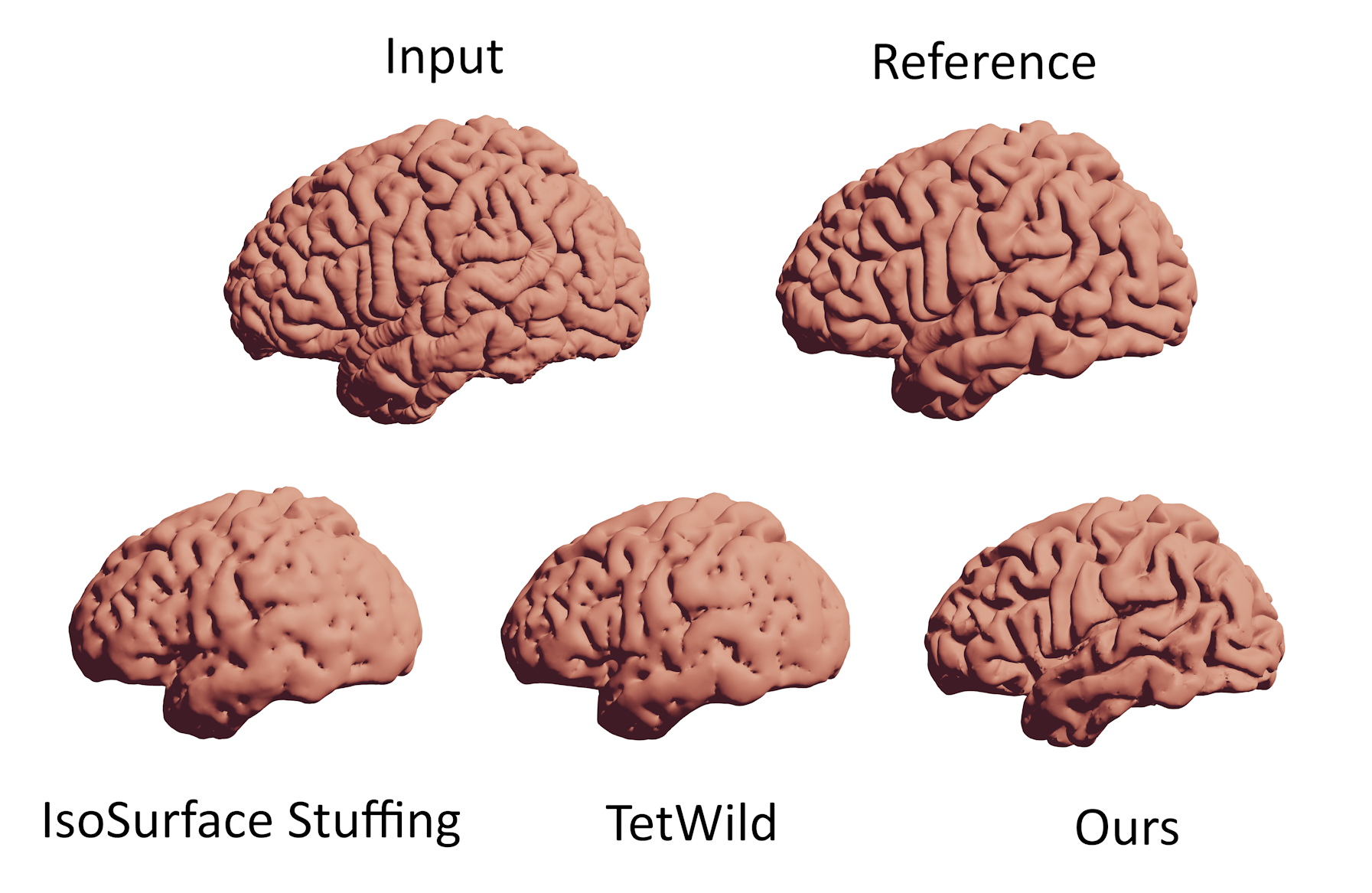}
    \caption{Comparison of smoothing simulation applied to tetrahedral meshes, highlighting connectivity issues. Top left: Input surface. Top right: Input surface after iterative smoothing. Bottom row: Results of smoothing applied to meshes generated by Isosurface Stuffing (left), TetWild (center), and our method (right).}
    \label{fig-sim-results}
  \end{adjustwidth}
\end{figure*}
To highlight the impact of different tetrahedralizations, we simulate an iterative smoothing process applied to the vertices, where vertices move towards the average position of directly connected surface vertices. As this simulation can be performed on both surface and volumetric meshes using their surface representations, it allows for a direct comparison with the original surface serving as ground truth. \cref{fig-sim-results} shows the varying simulation results of the different meshing algorithms. When applied to the original input surface, which possesses the correct surface connectivity, the smoothing process causes folds to become more separated. In contrast, for meshes generated by Isosurface Stuffing and TetWild, erroneous surface connections introduced during tetrahedralization prevent the folds from separating correctly. Instead, these connections artificially pull the folds towards one another. Our tetrahedralization method preserves the essential surface topology, resulting in deformation behavior similar to that observed on the reference input surface.

\newpage

\section{Discussion}
This work introduces a novel tetrahedralization method specifically designed for surfaces with self-intersections, such as cortical surfaces. A key strength of our approach is its ability to accurately preserve the surface connectivity of the input mesh, a property crucial for downstream applications like neuroimaging volumetric simulations where maintaining the separation between distinct anatomical regions is essential. We validated this capability using quantitative metrics. Our algorithm's success in maintaining the separation between nearby folds offers clear advantages over existing methods, particularly for deformation simulations. \\

While rigorous performance benchmarking is challenging due to variations in implementation environments (e.g., containerized vs. native, single- vs. multi-threaded), for the meshes used in this paper, both our algorithm and Isosurface Stuffing showed significantly faster execution times than TetWild for comparable element counts. Our approach prioritizes topological accuracy; for instance, it intentionally preserves distinct vertices even if they share the same geometric coordinates (twin points). This design choice ensures the accurate representation of the input surface's connectivity, preventing the artificial merging of regions that are geometrically close but topologically separate. If applications require unique vertex positions, post-processing steps such as minimal epsilon perturbations or merging twin points based on connectivity distance can be applied. \\

There is one notable area that can be improved. The mesh structure in regions of near-overlap can sometimes exhibit a jagged appearance, particularly when using fewer internal points. An edge-fixing post-processing step could resolve this issue. For example, in 2D, tracing shortest paths along interior nodes and then clamping edges onto original surface edges would increase the accuracy of edges near the boundary. However, robustly extending this to 3D presents edge cases and performance considerations that fell outside the scope of this project. \\

Beyond applications in neuroimaging, our method could be valuable for other domains requiring robust mesh generation from imperfect or noisy surface data where connectivity must be preserved. Future work could focus on further automating defect detection and mesh repair, optimizing for computational efficiency, and extending the technique for dynamic or multi-modal anatomical surfaces. Additionally, integrating this approach into existing neuroimaging pipelines could facilitate broader adoption and enable more accurate simulations of cortical mechanics.

\section{Conclusion}
In conclusion, we have presented a robust tetrahedralization method that excels at preserving the surface connectivity of cortical surfaces, even in the presence of self-intersections and near-touching folds. Compared to existing techniques, our approach offers distinct advantages for applications in neuroimaging and simulation that require accurate topological integrity and the separation of anatomical regions. Ultimately, this method can facilitate more realistic biomechanical simulations and anatomical analyses for neuroimaging.

\printbibliography

\end{document}